\documentclass[a4paper,fleqn,usenatbib]{mnras}

\usepackage{microtype}
\usepackage{graphicx}
\usepackage{booktabs}
\usepackage{newtxtext,newtxmath}
\usepackage[T1]{fontenc}
\usepackage{CJK}

\newcommand{\Msun}{\,\mathrm{M}_\odot}

\newcommand{\Rsun}{\,\mathrm{R}_\odot}

\newcommand{\Lsun}{\,\mathrm{L}_\odot}
\newcommand{\uHz}{\,\mu\mathrm{Hz}}

\newcommand{\dd}{\,\mathrm{d}}

\newcommand{\Gyr}{\,\mathrm{Gyr}}
\newcommand{\K}{\,\mathrm{K}}

\newcommand{\nm}{\,\mathrm{nm}}

\newcommand{\FeH}{[\mathrm{Fe}/\mathrm{H}]}
\newcommand{\Teff}{T_\mathrm{eff}}
\newcommand{\numax}{\nu_\mathrm{max}}
\newcommand{\eq}[1]{\begin{equation} #1 \end{equation}}
\newcommand{\st}[1]{_\mathrm{#1}}

\title[TESS observations of 12 Bo{\"o}tis]
  {Solar-like oscillations and ellipsoidal variations in TESS observations of the binary 12 Bo{\"o}tis}

\author[W.~H.~Ball et al.]{Warrick H. Ball,$^{1}$
  Andrea Miglio,$^{1,2,3}$
  William J. Chaplin,$^{1}$
  Keivan G. Stassun,$^{4}$
  \newauthor
  Rafael Garc{\'i}a,$^{5}$
  Lucia Gonz{\'a}lez-Cuesta,$^{6,7}$
  Savita Mathur,$^{6,7}$
  Thierry Appourchaux,$^{8}$
  \newauthor
  Othman Benomar,$^{9,10,11}$
  Derek L. Buzasi,$^{12}$
  Chen Jiang (姜晨),$^{13}$
  Cenk Kayhan,$^{14}$
  \newauthor
  Sibel {\"O}rtel,$^{15}$
  Zeynep~{\c C}elik~Orhan,$^{15}$
  Mutlu Y{\i}ld{\i}z,$^{15}$
  J.~M.~Joel Ong (王加冕),$^{16}$
  Sarbani Basu$^{16}$
  \\
  $^{1}$School of Physics and Astronomy, University of Birmingham, Edgbaston, Birmingham B15 2TT, United Kingdom\\
  $^{2}$Dipartimento di Fisica e Astronomia, Universit{\'a} degli Studi di Bologna, Via Gobetti 93/2, I-40129 Bologna, Italy\\
  $^{3}$INAF - Osservatorio di Astrofisica e Scienza dello Spazio di Bologna, Via Gobetti 93/3, I-40129 Bologna, Italy\\
  $^{4}$Department of Physics \& Astronomy, Vanderbilt University, Nashville, TN 37235, USA\\
  $^{5}$D\'epartement d'Astrophysique/AIM, CEA/IRFU, CNRS/INSU, Univ. Paris-Saclay \& Univ. de Paris, 91191 Gif-sur-Yvette, France\\
  $^{6}$Instituto de Astrof{\'i}sica de Canarias, La Laguna, Tenerife, Spain\\
  $^{7}$Dpto. de Astrof{\'i}sica, Universidad de La Laguna, La Laguna, Tenerife, Spain\\
  $^{8}$Universit{\'e} Paris-Sud, Institut d'Astrophysique Spatiale, UMR 8617, CNRS, B{\^a}timent 121, 91405 Orsay Cedex, France\\
  $^{9}$ Department of Astronomical Science, School of Physical Sciences, SOKENDAI, 2-21-1 Osawa, Mitaka, Tokyo 181-8588, Japan\\
  $^{10}$National Astronomical Observatory of Japan, 2-21-1 Osawa, Mitaka, Tokyo 181-8588, Japan\\
  $^{11}$Center for Space Science, New York University Abu Dhabi, P.O. Box 129188, Abu Dhabi, UAE \\
  $^{12}$Department of Chemistry and Physics, Florida Gulf Coast University, 10501 FGCU Blvd., Fort Myers, FL 33965 USA\\
  $^{13}$Max-Planck-Institut f{\"u}r Sonnensystemforschung, Justus-von-Liebig-Weg 3, 37077 G{\"o}ttingen, Germany\\
  $^{14}$Department of Astronomy and Space Sciences, Science Faculty, Erciyes University, 38030 Melikgazi, Kayseri, Turkey\\
  $^{15}$Department of Astronomy and Space Sciences, Science Faculty, Ege University, 35100, Bornova, {\.I}zmir, Turkey\\
  $^{16}$Department of Astronomy, Yale University, P.O. Box 208101, New Haven, CT 06520-8101, USA\\
}

\date{Accepted 2022 August 01. Received 2022 July 29; in original form 2022 May 23}
\pubyear{2022}

\begin{document}
\label{firstpage}
\pagerange{\pageref{firstpage}--\pageref{lastpage}}
\begin{CJK*}{UTF8}{gbsn}
\maketitle

\begin{abstract}
  Binary stars in which oscillations can be studied in either or
  both components can provide powerful constraints on our
  understanding of stellar physics.  The bright binary 12 Bo{\"o}tis (12 Boo)
  is a particularly promising system because the primary is roughly 60
  per cent brighter than the secondary despite being only a few per cent more massive.
  Both stars have substantial surface convection zones and
  are therefore, presumably, solar-like oscillators.  We report here
  the first detection of solar-like oscillations and ellipsoidal variations
  in the TESS light curve of 12 Boo.  Though the solar-like
  oscillations are not clear enough to unambiguously measure
  individual mode frequencies, we combine global asteroseismic
  parameters and a precise fit to the spectral energy
  distribution (SED) to provide new constraints on the properties of the
  system that are several times more precise than values in the literature.
  The SED fit alone provides new effective temperatures, luminosities
  and radii of $6115\pm45\K$, $7.531\pm0.110\Lsun$ and $2.450\pm0.045\Rsun$
  for 12 Boo A and $6200\pm60\K$, $4.692\pm0.095\Lsun$ and $1.901\pm0.045\Rsun$
  for 12 Boo B.  When combined with our asteroseismic constraints on 12 Boo A,
  we obtain an age of $2.67^{+0.12}_{-0.16}\Gyr$, which is consistent with
  that of 12 Boo B.
\end{abstract}

\begin{keywords}
  stars: oscillations (including pulsations); stars: individual (12 Boo); stars: binaries; asteroseismology
\end{keywords}

\section{Introduction}

Binary stars have long provided important tests of widely-used
one-dimensional stellar models.  Great attention is usually given
to double-lined eclipsing binaries, in which masses and radii can be measured, but astrometric
double-lined binaries also provide stellar masses, which are
arguably stars' most important initial physical parameter.

Independently, the study of stellar
oscillations---\emph{asteroseismology}---also provides important tests
of stellar physics.  When we can identify multiple modes and measure
their frequencies, each mode provides a slightly different average of
some interior properties, which allows very precise measurements of
certain characteristics (e.g., the mean density) and potentially
tests to distinguish between theories of the stars' interior physics
\citep[see e.g.][for a recent review]{aerts2021}.  In cool stars like
the Sun ($\Teff\lesssim6500\K$), near-surface convection drives and
damps oscillations across a wide range of frequencies.  These
oscillations are known as \emph{solar-like oscillations}; stars
that show these oscillations are \emph{solar-like oscillators}.  The
modes in solar-like oscillators can easily be identified by the
regular patterns they follow \citep[see e.g.][for a recent
  review]{garcia2019}.  Their study has recently been
revolutionised by space-based photometry from CoRoT
\citep{corot, corot2016}, \emph{Kepler} \citep{kepler} and K2 \citep{k2}.

The asteroseismology of binary stars is thus particularly promising
for testing stellar models but solar-like oscillations have been
measured in very few main-sequence or early subgiant binary systems.
\citet{miglio2014} predicted that \emph{Kepler} would detect
solar-like oscillations in both components of only a few main-sequence binaries,
which has been borne out.  Main-sequence binaries in which
\emph{Kepler} observed solar-like oscillations in both components
include 16 Cyg \citep[KIC 12069424, KIC 12069449;][]{metcalfe2012,davies2015}, HD 177412
\citep[KIC 7510397;][]{appourchaux2015} and HD 176465 \citep[KIC 10124866;][]{white2017}.
\citet{halbwachs1986} identified HIP 92961 \& 92962 (KIC 9139151 \& KIC 9139163)
as a pair with common proper motions and \citet{legacy2} included
both stars in their survey of solar-like oscillators observed by \emph{Kepler}.
Aside from these main-sequence binaries observed by \emph{Kepler}, ground-based
radial velocity campaigns have also observed solar-like oscillations
in both components of $\alpha$~Cen \citep{bouchy2001,kjeldsen2005}.

With the launch of the Transiting Exoplanet Survey Satellite
\citep[TESS;][]{tess}, we can now reverse the selection process and
search for solar-like oscillations in known main-sequence or subgiant
binaries.  TESS is less sensitive than \emph{Kepler}, so detecting
solar-like oscillations in main-sequence and subgiant stars is limited
to very bright stars with $G\lesssim6$ \citep[see, e.g., the
  non-detection of solar-like oscillations in AI Phe,][]{maxted2020}.
For example, solar-like oscillations have been detected
  in TESS observations of
  HD~221416 \citep{huber2019},
  $\nu$ Ind \citep{chaplin2020},
  $\lambda^2$ For \citep{nielsen2020}
  94 Aqr Aa \citep{metcalfe2020},
  HD~38529 \citep{ball2020},
  HD~19916 \citep{addison2021},
  $\rho$ CrB \citep{metcalfe2021},
  $\alpha$ Men A \citep{chontos2021},
  $\gamma$ Pav, $\zeta$ Tuc and $\pi$ Men \citep{huber2022}.

We report here the results of our search for solar-like oscillations
in the TESS light curve of
12 Bo{\"o}tis (d Bo{\"o}tis, HR 5304, HD 123999, TIC 418010485; hereafter 12 Boo),
a bright ($G=4.66$) binary system comprising two roughly $1.4\Msun$ stars
in a slightly eccentric ($e\approx0.2$), $9.6\dd$ orbit.
\citet{campbell1900} originally discovered the radial velocity variations in 12 Boo
and \citet{harper1914} subsequently measured more radial velocities
and fit orbital parameters.
\citet{merrill1922} conducted an interferometric survey of a number of binary stars
but could not resolve the orbit of 12 Boo.
The binary nature of the system was not revisited until \citet{abt1976}
computed new orbital elements.
\citet{demedeiros1999} measured new radial velocities as part of a survey
of evolved stars.

\citet{boden2000} presented the first interferometric observations of
the orbit of the system and noted that the primary is about $0.5$
magnitudes brighter in $V$ than the secondary despite being of about
the same mass.  They recognised that this probably meant the primary
has recently left the main-sequence and the secondary is near the end
of its main-sequence life.  \citet{boden2005} refined their previous
result and \citet{tomkin2006} independently derived consistent orbital
parameters.  \citet{konacki2010} presented an even more precise
simultaneous fit of the interferometric measurements combined
with the radial velocities they measured as part of the TATOOINE
search for circumbinary planets.  More recently, \citet{behr2011}
reported more independent radial velocities and \citet{kervella2017}
made a limited set of measurements of 12 Boo as a calibrator for their
study of $\alpha$ Cen.

\citet{miglio2007} highlighted potential of 12 Boo as a test for
stellar models.  In particular, they investigated how the detection of
mixed modes, which have oscillating components in the core and
envelope, would distinguish different possibilities for the extent of
extra mixing at the boundary of the main-sequence convective core.
\citet{miglio2007} also pointed out that even the more modest goal of
measuring the stars' large frequency separations $\Delta\nu$ would
provide more precise radius estimates.

We present here our analysis of TESS light curve of 12 Boo,
which includes the first detection of solar-like oscillations
and ellipsoidal variations in this system.  In Sec.~\ref{s:obs}, we detail our
new analyses of the TESS light curve and provide an updated fit of the
spectral energy distribution incorporating new photometry and parallax
data from Gaia.  In Sec.~\ref{s:mod}, we summarise our results from
fitting multiple sets of stellar models to the new constraints,
including asteroseismic constraints on 12 Boo A.  We close in
Sec.~\ref{s:outro} with some speculation on the future seismic
potential of 12 Boo.

\section{Observations}
\label{s:obs}

\begin{figure}
  \includegraphics[width=\columnwidth]{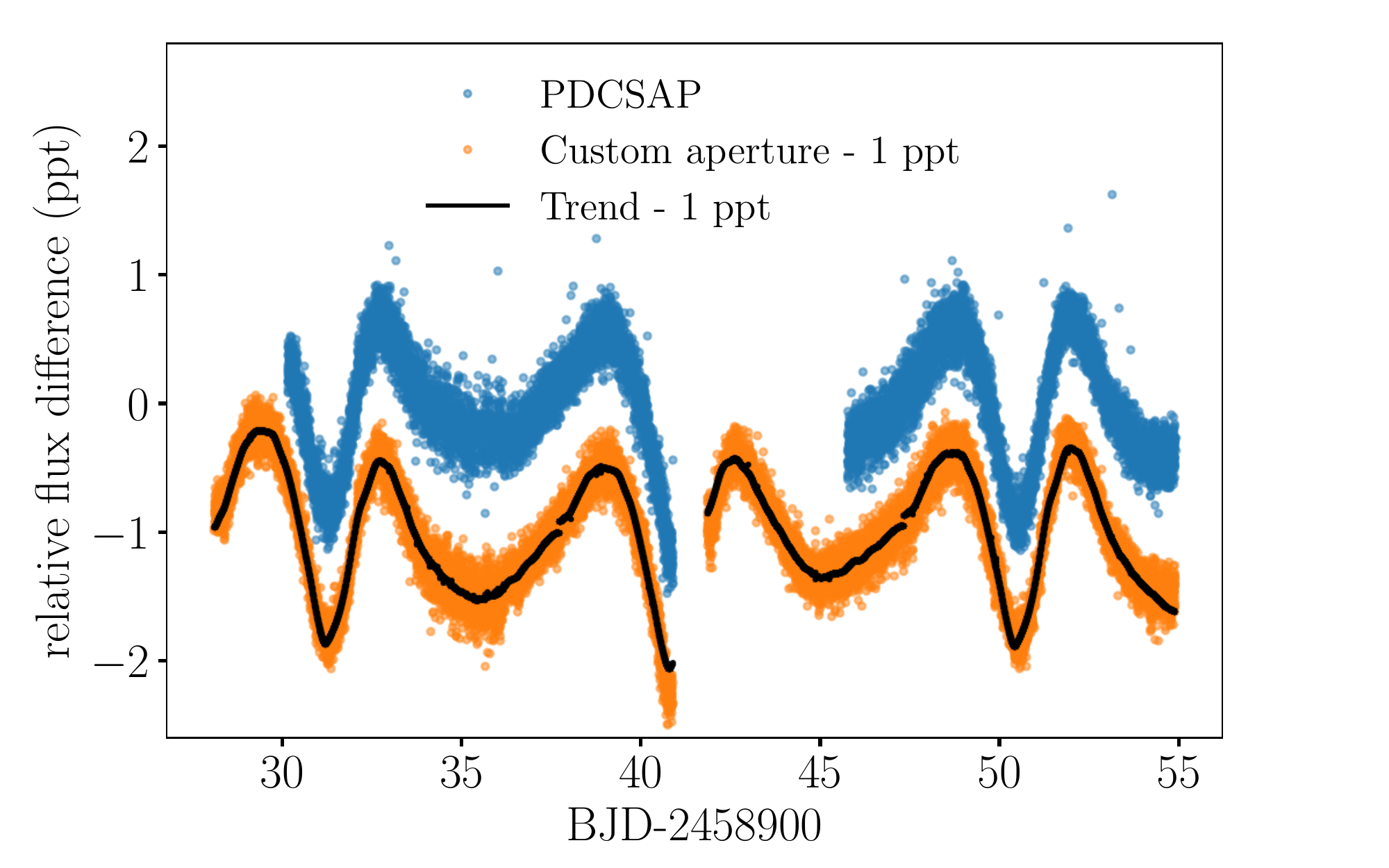}
  \caption{TESS light curves of 12 Boo from the PDCSAP pipeline (blue) and
    our custom reduction (orange, offset).  We subtracted a
    slowly-varying trend (black) before computing the power spectrum
    (Fig.~\ref{f:ps}).}
  \label{f:lc}
\end{figure}

\subsection{Light curve}

TESS observed 12 Boo on Camera 2 during Sector 23 (2020-03-19 to
2020-04-15) at a cadence of 2 min (i.e., TESS's original short cadence).  Our
initial inspection of the pipeline-reduced aperture photometry
(Pre-search Data Conditioning Simple Aperture Photometry, PDCSAP),
shown by the blue points in Fig.~\ref{f:lc}, immediately
showed mmag-level variations matching the orbital period of the system.
These are ellipsoidal variations caused by the gravity of each star slightly
distorting its companion.  As far as we know, this
is the first time ellipsoidal variations have been reported for 12
Boo.

The PDCSAP light curve, shown in blue in Fig.~\ref{f:lc}, offered only a marginal detection
of solar-like oscillations in 12 Boo A, so we experimented with custom
light curves, reduced from the short-cadence imagettes.
The imagettes include many cadences
that were excluded from the PDCSAP reduction
when scattered light left too little data to derive
cotrending basis vectors, which the pipeline uses to remove
systematic effects \citep[][quality bit 16]{tess_drn27}.
We selected all pixels with a median flux greater
than $100\mathrm{e}^-\,\mathrm{s}^{-1}$, including one line of pixels
above and below the saturated columns as well as the columns before
and after the saturated ones.
We then removed all data points taken while the
spacecraft was in safe mode (quality bit 2),
pointing to Earth (quality bit 4)
or desaturating the reaction wheels (quality bit 32).
We also removed data points marked as impulsive outliers (quality bit 10).
Finally, small gaps were filled using the inpainting techniques described by
\citet{garcia2014} and \citet{pires2015}.
Our custom light curve is shown
in orange in Fig.~\ref{f:lc}, offset downwards by $1\,\mathrm{ppt}$ for visibility.
The ellipsoidal variations remain clear and we recovered several days of data
near the beginnings of TESS's orbits.  These are the data that the PDCSAP
pipeline excluded because the scattered light prevents it from deriving
cotrending basis vectors.

To produce the seismic light curve, we applied a triangular filter with
a window function of 5 days ($2.31\,\uHz$).  To reduce the border
effects at the beginning and the end of the series, the filter was computed for the times covered
  in the original light curve after reflecting the original light
curve at each end by 2.5 days.  The trend is shown by the
black line in Fig.~\ref{f:lc} and effectively removes the ellipsoidal
variations, as well as slow systematic drifts at the beginnings and
ends of each of TESS's orbits.

\begin{figure}
  \includegraphics[width=\columnwidth]{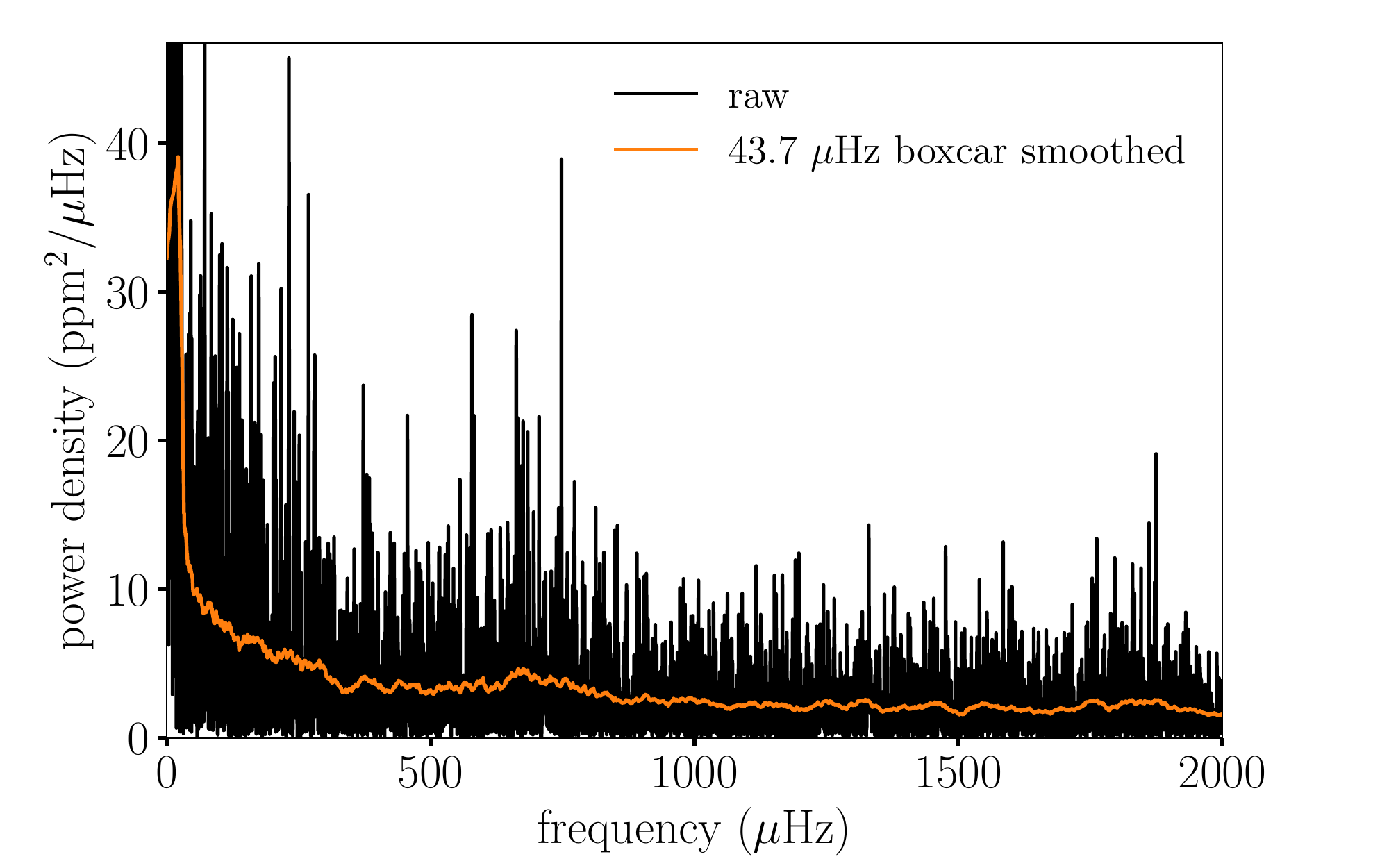}
  \caption{Raw (black) and smoothed (orange) power spectrum of our
    custom light curve for 12 Boo, computed using a Lomb--Scargle
    periodogram after subtracting the slowly-varying trend
    (Fig.~\ref{f:lc}).}
  \label{f:ps}
\end{figure}

\begin{figure}
  \includegraphics[width=\columnwidth]{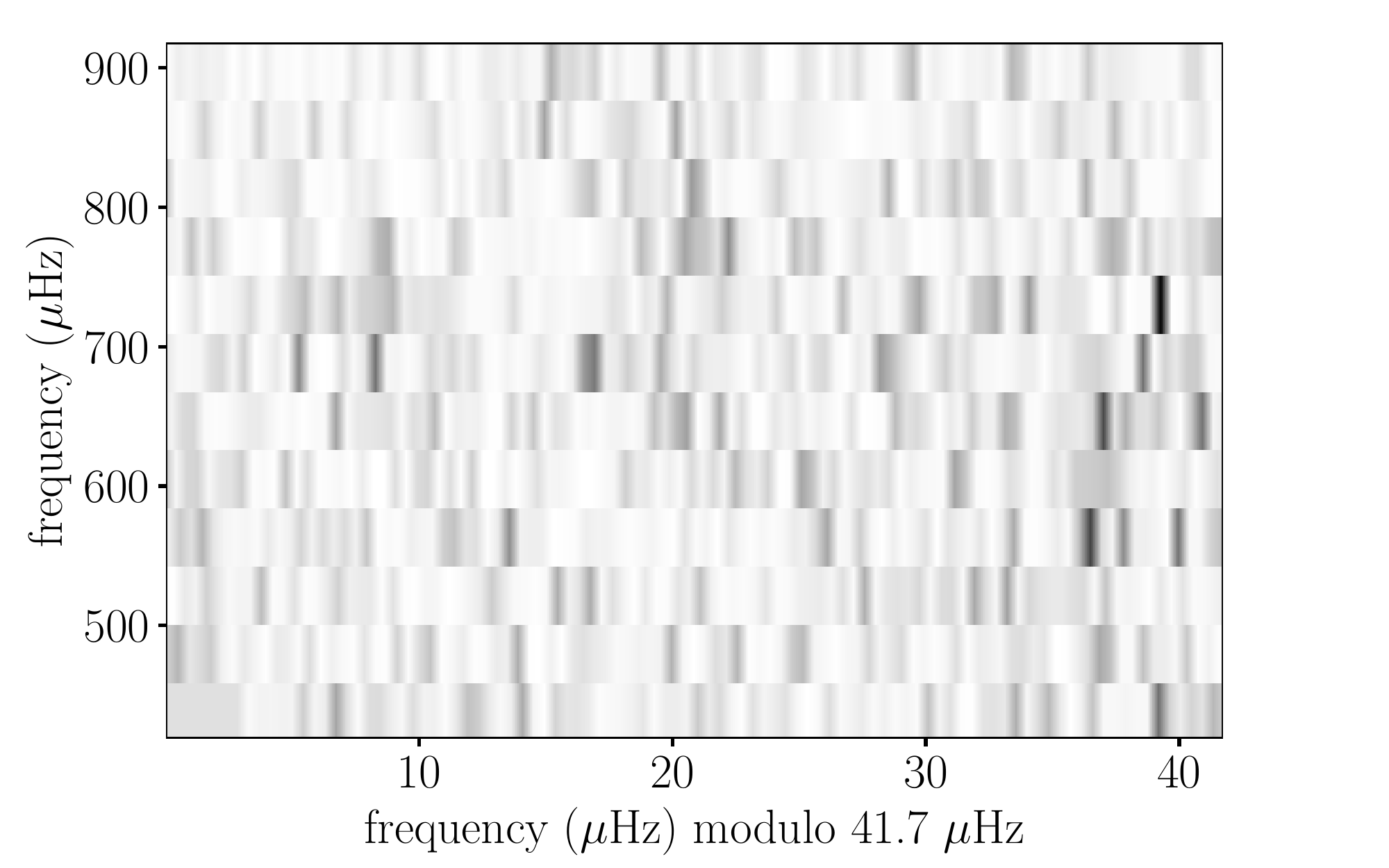}
  \caption{{\'E}chelle diagram of the raw power spectrum (c.f.~Fig.~\ref{f:ps}),
    folded on an approximate large separation of $\Delta\nu=41.7\uHz$.}
  \label{f:echelle}
\end{figure}

\begin{table}
  \begin{center}
    \begin{tabular}{lll}
      \toprule
      Team & $\Delta\nu/\uHz$ & $\numax/\uHz$ \\
      \midrule
      Benomar & $42.21_{-0.31}^{+0.38}$ & $696^{+17}_{-16}$ \\
      Jiang   & $41.6\pm0.5$            & $667\pm5$ \\
      Mathur  & $41.26\pm1.47$          & $669\pm40$ \\
      \midrule
      Combined & $41.7\pm1.0$ & $678\pm29$ \\
      \bottomrule
    \end{tabular}
  \end{center}
  \caption{Large frequency separations $\Delta\nu$ and frequencies of
    maximum oscillation power $\numax$ measured by three independent
    teams.}
  \label{t:seismo}
\end{table}

\begin{table}
  \renewcommand{\arraystretch}{1.2}
  \begin{tabular}{cccc}
    \toprule
    & 12 Boo A & 12 Boo B & Source \\
    \midrule
    & \multicolumn{3}{c}{Literature values} \\
    \midrule
    $\Teff/K$ & $6130\pm100$ & $6230\pm150$ & \citet{boden2005} \\
    $R/\Rsun$ & $2.474\pm0.095$ & $1.86\pm0.15$ &\citet{boden2005}  \\
    $L/\Lsun$ & $7.76\pm0.35$ & $4.69\pm0.74$ & \citet{boden2005} \\
    \midrule
    & \multicolumn{3}{c}{Constraints} \\
    \midrule
    $M/\Msun$ & $1.4109\pm0.0028$ & $1.3677\pm0.0028$ & \citet{konacki2010} \\
    $\Teff/K$ & $6115\pm45$ & $6200\pm60$ & Sec.~\ref{ss:sed} \\
    $R/\Rsun$ & $2.450\pm0.045$ & $1.901\pm0.045$ & Sec.~\ref{ss:sed} \\
    $L/\Lsun$ & $7.531\pm0.110$ & $4.692\pm0.095$ & Sec.~\ref{ss:sed} \\
    $\FeH$ & $-0.065\pm0.101$ & $-0.065\pm0.101$ & Sec.~\ref{ss:existing} \\
    \midrule
    & \multicolumn{3}{c}{Stellar modelling results} \\
    \midrule
    $R/\Rsun$ & $2.464^{+0.042}_{-0.025}$ & $1.881^{+0.056}_{-0.040}$ & Sec.~\ref{s:mod} \\
    $\bar\rho/(\mathrm{g\thinspace cm}^{-3})$ & $0.133^{+0.004}_{-0.006}$ & $0.290^{+0.019}_{-0.026}$ & Sec.~\ref{s:mod} \\
    $\log g$ & $3.802^{+0.010}_{-0.014}$ & $4.028^{+0.023}_{-0.028}$ & Sec.~\ref{s:mod} \\
    $t/\Gyr$ & $2.67^{+0.12}_{-0.16}$ & $2.66^{+0.11}_{-0.15}$ & Sec.~\ref{s:mod} \\
    \bottomrule
  \end{tabular}
  \caption{Various properties of 12 Boo.  The upper set include
    constraints in the stellar modelling (Sec.~\ref{s:mod}) and the
    lower set the results of the modelling.}
  \label{t:properties}
\end{table}

\subsection{Asteroseismic parameters}
\label{ss:astero}

Fig.~\ref{f:ps} shows the power spectrum of our detrended custom light
curve after subtracting the trend determined above.  There is a clear excess of power around $700\uHz$, 
where the oscillations of the primary are expected based on its previously measured
mass, luminosity and radius (see Table~\ref{t:properties})
combined with the $\nu\st{max}$ scaling relation \citep{brown1991}.
There is no clear excess for the secondary, whose oscillations should peak
roughly around $1200\uHz$ based on a similar calculation using its properties
from the literature.

Fig.~\ref{f:echelle} shows the echelle diagram around the obvious power excess,
folded on a large separation $\Delta\nu=41.7\uHz$.
This produces a roughly vertical ridge, which is characteristic
of solar-like oscillations, around $40\uHz$.  Three teams
\citep{benomar2012,jiang2011,mathur2010} independently analysed the power
spectrum to measure the properties of the oscillations.
The teams reported mutually consistent values for
the frequency of maximum oscillation power $\numax$
and large separation $\Delta\nu$, shown in Table~\ref{t:seismo}.

The teams did not, however, reach a consensus on the identity of any
individual mode frequencies, at least partly because the properties of 12 Boo
A place it where empirical relations for identifying
the angular degrees, which use their horizontal offset in the
  echelle diagram $\epsilon$, are uncertain \citep[e.g.][]{white2012}.
The identification is further confounded by the potentially mixed modes,
which would deviate from the simple asymptotic spacing of high-order pressure modes,
and the rotational splitting, which would be around $1.2\uHz$ if 12 Boo A's
rotation is aligned and synchronised with its orbit.
\citet{boden2000} noted that, if the rotation axes are aligned with the orbit,
the rotation periods are slightly shorter than the orbital
period, presumably because the tidal torques are strongest
at periastron, when the stars are moving relatively quickly
\citep[see e.g.][]{hut1981}.

We therefore proceeded to analyse 12 Boo A using $\numax$ and
$\Delta\nu$ but no individual frequencies.  Our combined mean values
for $\numax$ and $\Delta\nu$ are the means of the three estimates.
Our consolidated variance is the mean of the variances plus the
variance of the means.  We symmetrized the slightly
asymmetric results by Benomar by taking the mean of the $\pm1\sigma$ limits as
the central value.  This gives $\numax=678\pm29\uHz$ and
$\Delta\nu=41.7\pm1.0\uHz$, as shown in Table~\ref{t:seismo}.

\begin{figure}
  \includegraphics[width=\columnwidth]{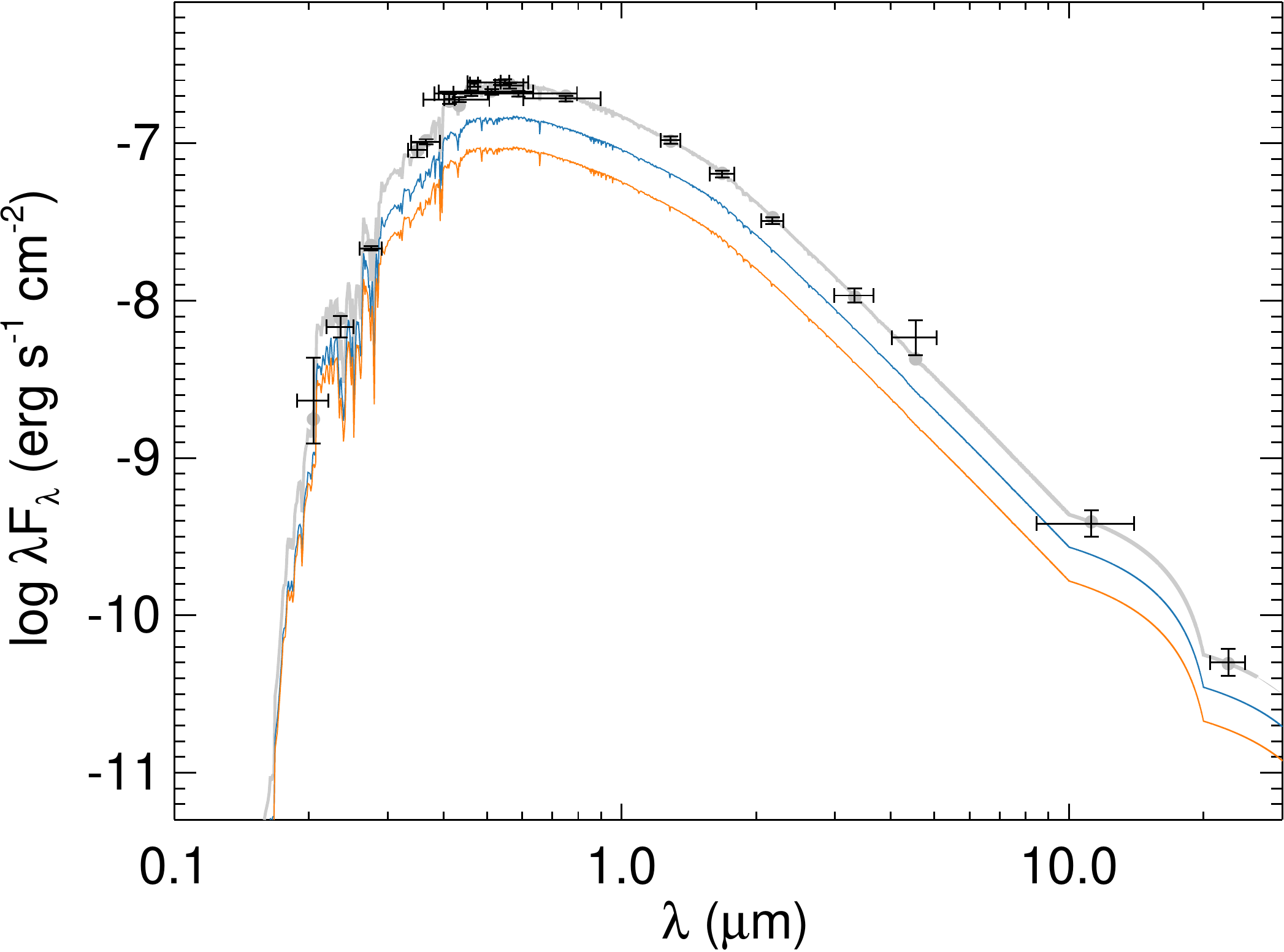}
  \caption{Spectral energy distribution (SED) of 12 Boo.  Each black cross
    represents an observed passband flux at its central wavelength.  The vertical
    bar indicates the measured uncertainty and the horizontal bar the
    wavelength range of the passband.
    The overall SED is shown in grey, with additional grey points showing the
    model value for each passband.  The blue and orange
    curves show the SEDs of the primary and secondary, respectively.}
  \label{f:sed}
\end{figure}

\begin{table}
  \begin{center}
  \begin{tabular}{rrrrr}
                &    $R_A$ &    $R_B$ & ${\Teff}_{,A}$ & ${\Teff}_{,B}$ \\
          $R_A$ & $ 1.000$ & $ 0.801$ &    $-0.930$ &    $-0.656$ \\
          $R_B$ & $ 0.801$ & $ 1.000$ &    $-0.701$ &    $-0.909$ \\
    ${\Teff}_A$ & $-0.930$ & $-0.701$ &    $ 1.000$ &    $ 0.562$ \\
    ${\Teff}_B$ & $-0.656$ & $-0.909$ &    $ 0.562$ &    $ 1.000$ \\
  \end{tabular}
  \end{center}
  \caption{Correlation coefficients in SED fit.}
  \label{t:corr}
\end{table}

\subsection{Spectral energy distribution}
\label{ss:sed}

Reported luminosities of 12 Boo A and B pre-date the precise
magnitudes and parallaxes from Gaia.  We therefore computed a new fit
to the total spectral energy distribution (SED) to determine the
individual luminosities and effective temperatures, using the methods
described by \citet{stassun2016} and \citet{stassun2017,stassun2018}.
The photometry comprises 
fluxes in the $197\nm$, $237\nm$ and $274\nm$ (ultraviolet) passbands of TD-1 \citep{td1},
$UBV$ magnitudes from \citet{mermilliod2006},
$B\st{T}V\st{T}$ magnitudes from Tycho-2 \citep{tycho2a,tycho2b}
Str{\"o}mgren $uvby$ magnitudes from \citet{paunzen2015},
$JHK\st{S}$ magnitudes from 2MASS \citep{2mass},
$W1$--$4$ magnitudes from WISE \citep{wise},
and Gaia $G$, $G\st{BP}$ and $G\st{RP}$ magnitudes from Early Data Release 3
\citep[EDR3;][]{gaia,gaia_edr3,riello2021}.
The two stars were distinguished by using flux ratios derived
from the $V$-band magnitude difference in \citet{boden2005}
and the $H$-band magnitude difference in \citet{kervella2017}.

The best fitting radii $R$ and effective temperatures $\Teff$ are
shown in Table~\ref{t:properties}, along with the luminosities $L$
derived using the Stefan--Boltzmann law.  Fig.~\ref{f:sed} shows the
observed photometry along with the total SED and the SEDs of the two
stars.  Because both stars are fit simultaneously and are spectrally
so similar, the fit parameters are strongly correlated.  We have included the
correlation coefficients of the radii and effective temperatures in
Table~\ref{t:corr}.

\subsection{Existing complementary data}
\label{ss:existing}

We choose to use the masses measured by \citet{konacki2010}, whose radial velocities
have the smallest residuals (about $34.0$ and $38.3\,\mathrm{m}\,\mathrm{s}^{-1}$
for 12 Boo A and B) and whose interferometry has the greatest phase coverage.
The composition of a star---usually expressed through its metallicity $\FeH$---is an
important constraint on its evolution.  For 12 Boo, we use the
spectroscopic measurements by \citet{balachandran1990} and
\citet{lebre1999} of $-0.03\pm0.09$ and $-0.1\pm0.1$, respectively,
which we combine (as we did $\numax$ and $\Delta\nu$ above) to
obtain $\FeH=-0.065\pm0.101$.  We assume both stars have surface abundances
that are the same within this uncertainty. 

\section{Stellar modelling}
\label{s:mod}

Four teams computed best-fitting stellar models of the primary using a
number of established methods \citep{ball2020,jiang2021,celikorhan2021} that fit
predictions by various stellar models
\citep{astec,demarque2008,paxton2019} and stellar oscillation programs
\citep{antia1994,adipls,gyre1,gyre2} to the specified constraints.
Three of the teams also modelled the secondary.  As constraints for
both stars, they used the masses $M$, metallicities $\FeH$,
luminosities $L$ and effective temperatures $\Teff$ in
Table~\ref{t:properties}.  For 12 Boo A, they also used the
consolidated $\numax$ and $\Delta\nu$ from Table~\ref{t:seismo}.

We combined the results from each team using a linear opinion pool
\citep{stone1961},\footnote{Though this is the usual modern reference
for the idea, \cite{bacharach1979} attributes the concept to \citet{laplace1814}.}
where the overall probability distribution is taken as the equally-weighted
sum of the probability distribution from each modeller.
Some teams reported strongly asymmetric uncertainties, so we fit the results from
each team using the lognormal distribution defined by \citet[][Sec.~A.8]{hosking1997},
which allows arbitrary skewness and has the normal distribution as the limit when the shape parameter
is zero.  We then interpolated the total cumulative distribution function at the percentiles
that correspond to the median and $\pm1\sigma$ limits of a normal distribution,
and report these as our central results and (asymmetric) uncertainties.

The modelling results are also shown in Table~\ref{t:properties}.  We
note that the stars' ages are constrained to within about 6 per cent.
This is principally because the masses are very precisely known.  Ages
are often imprecise because of the correlation with other parameters,
including the mass.  Because main-sequence lifetimes $t\st{MS}$ scale roughly
like $M^{-3}$, one na{\"i}vely expects fractional age uncertainties
at least about $3$ times larger than the fraction mass uncertainties.
This precision on the ages is not typical of
asteroseismology of individual stars, whose masses are less tightly constrained.

\section{Discussion and conclusions}
\label{s:outro}

We have analysed the TESS light curve of the binary star 12 Boo, with
the aim of combining individual seismic frequencies with the precise
masses from the stars' observed mutual orbit.  We were unable to
robustly identify individual frequencies but have used new photometric
constraints, the parallax measurement from Gaia and the global oscillation
properties---$\numax$ and $\Delta\nu$---of 12 Boo A to revise the
properties of the system.  Crucially, this includes the radius, which is
otherwise only weakly constrained in this non-eclipsing system.

The SED fit alone significantly improves the precision of the
individual components' properties.  We compare our results to those of
\citet{boden2005}, who appear to have most recently evaluated the
total SED, and whose values are also reported in
Table~\ref{t:properties}.  The dramatic improvement is driven by the
precise photometry from Gaia and the additional differential $H$-band
magnitude by \citet{kervella2017}.  There is little improvement through the
Gaia EDR3 parallax of $27.484\pm0.117\,\mathrm{mas}$, which is only
slightly more precise than the orbital parallax of
$27.72\pm0.15\,\mathrm{mas}$ that \citet{boden2005} derived.  For 12
Boo B, \citet{boden2005} estimated a radius of $1.86\pm0.15\Rsun$,
compared to our $1.901\pm0.045\Rsun$ from the SED fit.  That is, our
new radius is three times as precise.  Our radius of
$1.881^{+0.056}_{-0.040}\Rsun$ from detailed stellar modelling
reflects this constraint.  For 12 Boo A, our SED fit gives a radius
of $2.450\pm0.045\Rsun$, which is about twice as precise as the
estimate of $2.474\pm0.095\Rsun$ given by \citet{boden2005}.

Correspondingly, we have improved the bolometric luminosity estimates
for 12 Boo A and B from $7.76\pm0.35\Lsun$ and $4.69\pm0.74\Lsun$
to $7.531\pm0.110\Lsun$ and $4.692\pm0.095\Lsun$.
The new luminosities are about three and seven times more precise.

The radius of 12 Boo A is further constrained by our measurements
of the asteroseismic parameters $\Delta\nu$ and $\numax$.  If we
simply use the scaling relation \citep{ulrich1986,kjeldsen1995}
\eq{\Delta\nu\propto\sqrt{\bar\rho}\propto\sqrt{\frac{M}{R^3}}}
we obtain a radius $2.456\pm0.039\Rsun$, which is slightly more precise
than the SED fit.  The stellar modelling result, which has asymmetric
uncertainties, is more precise still.
The mean density $\bar\rho$ and, to a lesser extent, surface gravity $\log g$ of 12
Boo A are much better constrained than 12 Boo B.  This is a natural
consequence of the seismic data for 12 Boo A: $\Delta\nu$ tightly constrains
$\bar\rho$; $\numax$ constrains $\log g$.

Our results thus significantly improve the properties of the system
that have been reported in the literature.
A more sophisticated analysis of the light curve and power spectrum
might allow the identification of individual mode frequencies.
Furthermore, 12 Boo is scheduled to be re-observed during TESS's Sector 50
(2022-03-26 to 2022-04-22), and additional data
at a cadence of 120 or 20 seconds might allow
for a mode identification through which the Sector 23 data can be
better exploited.

12 Boo might also be an interesting target for the Stellar Oscillation
Network Group \citep[SONG;][]{song,song2019}, which aims to become a
worldwide network of telescopes with which to measure radial velocity
variations in bright solar-like oscillators.  The signal will be complicated
by the stars' orbital motions but the scientific value of measuring
the individual mode frequencies remains, for now, unexplored.

\section*{Acknowledgements}

WHB and WJC thank the UK Science and Technology Facilities Council
(STFC) for support under grant ST/R0023297/1.
AM acknowledges support from the ERC Consolidator Grant funding scheme (project ASTEROCHRONOMETRY, \url{https://www.asterochronometry.eu}, G.A. n. 772293).
RAG acknowledges the support from PLATO and GOLF CNES grants.
LGC thanks the support from grant FPI-SO from the Spanish Ministry of
Economy and Competitiveness (MINECO) (research project
SEV-2015-0548-17-2 and predoctoral contract BES-2017-082610).
SM acknowledges support from the Spanish Ministry of Science and Innovation (MICINN) with the Ram\'on y Cajal fellowship no.~RYC-2015-17697 and grant no.~PID2019-107187GB-I00, and through AEI under the Severo Ochoa Centres of Excellence Programme
2020--2023 (CEX2019-000920-S).
DLB acknowledges support from the TESS GI Program under NASA awards 80NSSC18K1585 and 80NSSC19K0385.
JC is supported by a grant from the Max Planck Society to prepare for the scientific exploitation of the PLATO mission.
CK is supported by Erciyes University Scientific Research Projects
Coordination Unit under grant number DOSAP MAP-2020-9749.

This paper includes data collected by the TESS mission, which are
publicly available from the Mikulski Archive for Space Telescopes
(MAST).  Funding for the TESS mission is provided by the NASA Explorer
Program.

This work has made use of data from the European Space Agency (ESA) mission
{\it Gaia} (\url{https://www.cosmos.esa.int/gaia}), processed by the {\it Gaia}
Data Processing and Analysis Consortium (DPAC,
\url{https://www.cosmos.esa.int/web/gaia/dpac/consortium}). Funding for the DPAC
has been provided by national institutions, in particular the institutions
participating in the {\it Gaia} Multilateral Agreement.

\section*{Data availability}

Original TESS light curves and pixel-level data are available from the
Mikulski Archive for Space Telescopes at \url{http://mast.stsci.edu/}.
Other data underlying this article will be shared on reasonable
request to the corresponding author.

\bibliographystyle{mnras}
\bibliography{../master}

\end{CJK*}
\bsp	
\label{lastpage}
\end{document}